# A Deep Learning Algorithm Based on CNN-LSTM Framework for Predicting Cancer Drug Sales Volume

Yinghan Li [1,4], Yilin Yao[1,5], Junghua Lin [2,6], Nanxi Wang[3,7]

[1] School of International Business, Henan University, Henan, 450001, China
[2] Suffolk university, 02128, USA
[3] University of Southern California, Los Angeles, CA, 90089, USA

[4] 15239528393@163.com
[5] Yeechen1942590243@gmail.com
[6] rjl00774@su.suffolk.edu
[7] nanxiwan@usc.edu

**Abstract.** This study explores the application potential of a deep learning model based on the CNN-LSTM framework in forecasting the sales volume of cancer drugs, with a focus on modeling complex time series data. As advancements in medical technology and cancer treatment continue, the demand for oncology medications is steadily increasing. Accurate forecasting of cancer drug sales plays a critical role in optimizing production planning, supply chain management, and healthcare policy formulation. The dataset used in this research comprises quarterly sales records of a specific cancer drug in Egypt from 2015 to 2024, including multidimensional information such as date, drug type, pharmaceutical company, price, sales volume, effectiveness, and drug classification. To improve prediction accuracy, a hybrid deep learning model combining Convolutional Neural Networks (CNN) and Long Short-Term Memory (LSTM) networks is employed. The CNN component is responsible for extracting local temporal features from the sales data, while the LSTM component captures long-term dependencies and trends. Model performance is evaluated using two widely adopted metrics: Mean Squared Error (MSE) and Root Mean Squared Error (RMSE). The results demonstrate that the CNN-LSTM model performs well on the test set, achieving an MSE of 1.150 and an RMSE of 1.072, indicating its effectiveness in handling nonlinear and volatile sales data. This research provides theoretical and technical support for data-driven decision-making in pharmaceutical marketing and healthcare resource planning.

## 1. Introduction

With the global aging population accelerating, the incidence of cancer rising, and precision medicine advancing rapidly, the cancer drug market is undergoing unprecedented growth. As high-value and high-demand pharmaceutical products with complex supply chain dependencies, cancer drugs require precise demand forecasting to ensure efficient production planning, inventory management, and rational healthcare policy design [1]. In the post-pandemic era, where pharmaceutical supply chains face increasing instability, accurate forecasting of cancer drug sales has become a critical concern across medical, economic, and policy-making domains.

Cancer drug sales are influenced by a wide range of factors, including advancements in treatment protocols, drug pricing strategies, changes in insurance coverage, seasonal healthcare behaviors, disease prevalence, and pharmaceutical marketing efforts [2]. Traditional forecasting models, such as linear regression and classical time series approaches, often fail to capture nonlinear fluctuations, sudden market changes, and intricate interactions among multiple variables. With the growing availability of medical big data and the rapid development of artificial intelligence technologies, deep learning-based forecasting models offer a promising alternative to address these complexities.

This study constructs a forecasting model based on a hybrid deep learning framework that combines Convolutional Neural Networks (CNN) [3] and Long Short-Term Memory (LSTM) [4] networks to improve the accuracy of cancer drug sales prediction. The dataset used spans from 2015 to 2024 and includes quarterly sales records of a specific cancer drug in Egypt. It contains multidimensional attributes such as timestamp, drug type, pharmaceutical company, price, sales volume, efficacy rating, and drug classification. In the proposed framework, the CNN component is responsible for extracting local temporal features from the input sequences, while the LSTM component captures long-term dependencies and seasonal trends in the sales data, enabling robust modeling of complex patterns.

To evaluate the model's performance, two commonly used metrics—Mean Squared Error (MSE) [5] and Root Mean Squared Error (RMSE)—are employed. Experimental results demonstrate that the CNN-LSTM model achieves strong predictive performance, with a final MSE of 1.150 and RMSE of 1.072. These results confirm the model's effectiveness in capturing nonlinear dynamics and enhancing forecasting accuracy for high-dimensional and volatile medical sales data. This research provides a data-driven approach to support pharmaceutical industry stakeholders in production optimization and strategic planning and contributes valuable insights to the broader field of intelligent healthcare forecasting.

## 2. Literature Review

Time series forecasting of cancer drug sales plays a vital role in medical resource allocation, pharmaceutical inventory management, production scheduling optimization, and healthcare policy formulation. With the continuous rise in cancer incidence and the advancement of precision medicine, cancer drugs—being high-value and highly supply-dependent medical products—exhibit complex demand patterns characterized by volatility and seasonality. Accurate forecasting of their sales trends is essential for ensuring timely drug availability for patients, reducing inventory costs, and improving the efficiency of pharmaceutical supply chains. However, due to the highly nonlinear, cyclical, and multi-factorial nature of cancer drug sales data, traditional forecasting methods still face significant challenges in terms of accuracy and generalization. In recent years, deep learning-based forecasting models have been rapidly developed, demonstrating remarkable advantages in handling high-dimensional and complex time series data, with notable improvements in prediction accuracy.

Konstantinos P. Fourkiotis et al [6]. conducted a study aiming to improve pharmaceutical sales predictions by combining conventional methods like ARIMA with advanced machine learning models such as LSTM neural networks and XGBoost. Through the analysis of a large dataset, their findings demonstrate the effectiveness of XGBoost in achieving more precise forecasts, indicating the value of hybrid approaches in this field. Noura Qassrawi et al [7]. explored Deep Neural Network algorithms to forecast drug sales and prices. They found Long Short-Term Memory outperformed MLP and CNN in predicting sales of new drugs with limited historical data.

Raman Pall et al [8]. constructed machine learning models using data from Canadian pharmacies to predict drug shortages. Their models achieved 69% accuracy in classifying shortage levels and anticipated 59% of the most impactful shortages a month in advance without manufacturer inventory data. This work enables pharmacists to optimize orders and mitigate drug shortage effects.

Semen Budennyy et al [9]. proposed a machine learning framework to predict pharma market reactions to clinical trial announcements. Their model uses BERT for sentiment extraction and Temporal Fusion Transformers for forecasting. Using the FDA dataset, the study found more pronounced reactions to negative clinical news and showed the framework's effectiveness with ROC AUC scores over 0.7.

## 3. Data Introduction

The dataset used in this study consists of quarterly sales records of a specific cancer drug in Egypt, spanning from 2015 to 2024 and covering a total of 40 quarters. The data were collected through web scraping from several reputable Egyptian pharmaceutical websites and online medical platforms, encompassing a wide range of cancer drugs available on the local market. Each record contains multidimensional structured information, including timestamp (quarter), drug type, pharmaceutical company, unit price, sales volume, effectiveness rating, and drug classification. This dataset provides a valuable foundation for analyzing market demand trends, sales dynamics, and pricing strategies in Egypt's pharmaceutical sector. It holds significant research potential for applications in market analysis, healthcare policy design, and drug supply chain optimization.

To ensure the quality and expressiveness of the input data, a systematic data preprocessing and cleaning process was applied before model training. First, all invalid entries containing missing values—represented by placeholders such as -99 or NaN in the original dataset—were removed to prevent distortion of the model's learning process. Next, categorical fields such as drug type, manufacturer, and classification were standardized and uniformly encoded to maintain semantic consistency across heterogeneous sources. In addition, unit normalization and time alignment were performed on numeric fields like price and sales volume to enhance the model's ability to capture long-term trends and local fluctuations.

Through this structured data processing pipeline, we constructed a well-organized, semantically enriched, and high-quality time series dataset of cancer drug sales. This dataset not only improves the stability and generalizability of model training but also provides a solid foundation for the CNN-LSTM-based deep learning algorithm applied in the forecasting task.

**Table 1.** Variables and descriptions

| variable | description |
|---|---|
| Drugname | The name of the medication. |
| Price | The price of the drug in Egyptian pounds (EGP). |
| Date | The timestamp when the data was scraped |
| Form | The form of the medication |
| Company | The company manufacturing the drug |
| Region | Placeholder for the region where the drug is commonly available |
| SalesVolume | The sales volume of this drug |
| Effectiveness | The effectiveness of anti-cancer drugs |
| UserEvaluate | User reviews of purchasing the drug on public websites |

Table 1 lists the variables included in the Cancer Drug Sales Volume Dataset for Egypt from 2015 to 2024. This dataset covers key attributes related to cancer drug sales in Egypt over a decade, on a quarterly basis. The variables encompass drug specifics such as "Drugname" (the name of the medication), "Form" (the form of the medication), and "Company" (the company manufacturing the drug). Pricing information is represented by "Price" (the price of the drug in Egyptian pounds).

Temporal data is provided by "Date" (the timestamp when the data was scraped). Sales data is indicated by "SalesVolume" (the sales volume of the drug). Effectiveness - related data is shown by "Effectiveness" (the effectiveness of anti - cancer drugs). User feedback is captured by "UserEvaluate" (user reviews of purchasing the drug on public websites). Additionally, there is a placeholder variable "Region" for the region where the drug is commonly available.

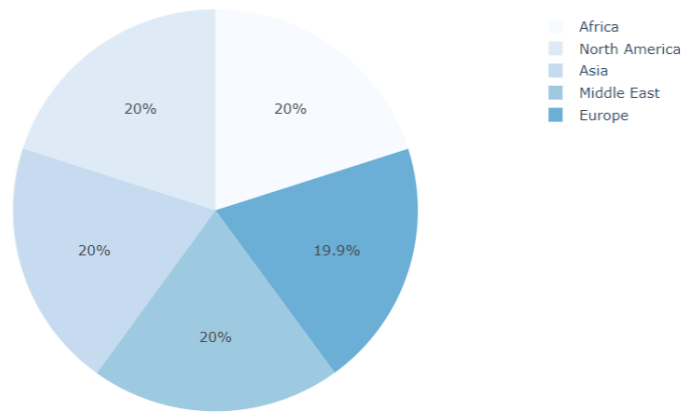

**Figure 1.** Pharmaceutical Company Regional Distribution

Figure 1 visually presents the geographical distribution of pharmaceutical companies within the Cancer Drug Sales Volume Dataset for Egypt from 2015 to 2024. The pie chart vividly illustrates the proportion of companies in different regions contributing to the dataset. As can be seen from the figure, the distribution is relatively even across the five major regions, with Africa, North America, Asia, and the Middle East each accounting for exactly 20%. Europe, on the other hand, holds a slightly smaller proportion at 19.9%, yet it remains a significant contributor to the dataset. This distribution implies a broad and diversified pharmaceutical market in Egypt, sourcing medications from various global regions.

**4. CNN-LSTM Model Introduction**
*4.1. CNN Layer*
In the CNN-LSTM deep learning framework proposed in this study, the Convolutional Neural Network (CNN) is employed to extract local features from the time series data of cancer drug sales, leveraging its advantages in processing grid-structured sequential data [10]. As a feedforward neural network, CNN excels at local perception and weight sharing, enabling it to effectively identify local patterns within the input data and thereby enhance the model's representational capacity and training efficiency in complex sequence modeling tasks.

For the structured time series data used in this study—which includes multiple features such as timestamps, drug types, pharmaceutical companies, prices, and sales volumes—the CNN component first transforms the raw input sequence into a multi-channel feature map [11]. Each channel corresponds to a specific feature (e.g., price, effectiveness rating), and one-dimensional convolutional kernels slide along the temporal axis to automatically learn the variations of local sales trends within different time windows. For instance, subtle price fluctuations across consecutive quarters, sudden increases or decreases in sales volume, and periodic patterns associated with specific drug categories can all be effectively captured through the convolutional operations.

In terms of architectural design, we adopted two layers of one-dimensional convolution with kernel sizes set to 3 and 5, respectively, to capture local temporal patterns at different scales. The number of convolutional filters in each layer is set to 64 and 128. Each convolutional layer is followed by Batch Normalization and a ReLU activation function to enhance the model's nonlinear representation capability and mitigate gradient vanishing issues. To further reduce the dimensionality of the feature maps and computational cost while retaining critical information, a max-pooling layer with a pool size of 2 is applied after each convolutional block. This structure significantly improves training efficiency and generalization, and helps prevent overfitting to redundant information.

Through this configuration, the CNN component is able to efficiently compress the input while preserving essential local temporal features of sales dynamics. These high-level features extracted by CNN are subsequently passed to the LSTM module, which further models long-term dependencies and global sequential patterns, forming a collaboratively optimized deep learning architecture.

*4.2. LSTM Layer*

The Long Short-Term Memory (LSTM) network serves as the core temporal modeling component in this study, designed to capture long-term dependencies and global trends within the cancer drug sales data [12]. LSTM is an improved type of Recurrent Neural Network (RNN) [13], incorporating gating mechanisms—namely, input gate, forget gate, and output gate—that effectively address the issues of vanishing or exploding gradients commonly encountered in traditional RNNs when dealing with long sequences. This makes LSTM particularly well-suited for modeling time series data with long-range dependencies.

In the context of cancer drug sales forecasting, fluctuations in sales volume are influenced not only by short-term factors such as quarterly price changes, supply-demand dynamics, and policy shifts, but also by long-term market trends, seasonal cycles, and ongoing treatment demands. Relying solely on local information is insufficient for accurate prediction. LSTM enables the retention and transfer of temporal state information, allowing the model to understand deep temporal correlations across quarters or even years, thereby enhancing the overall accuracy and robustness of the predictions.

In terms of implementation, we constructed a deep architecture consisting of two stacked LSTM layers built upon the local temporal features extracted by the CNN module. Each LSTM layer contains 128 hidden units, a configuration that balances expressive capacity with computational efficiency—particularly suitable for moderate-length time series at the quarterly level. To prevent overfitting, a dropout layer with a dropout rate of 0.3 is inserted between the LSTM layers to improve the model's generalization. Additionally, we adopted a stateful LSTM architecture with a state reset mechanism to preserve temporal dependencies across multiple training batches, thereby enhancing training stability.

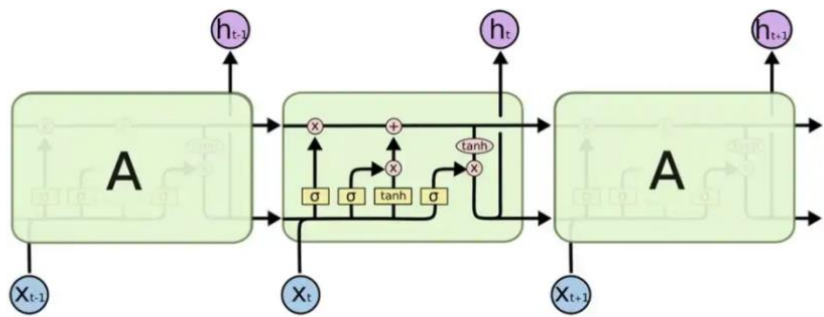

**Figure 2.** The structure of LSTM [14]

Finally, the high-level sequential representations produced by the LSTM layers are passed to a fully connected dense layer for regression output, yielding the predicted sales volume for the next quarter. This design enables the LSTM module to effectively model nonlinear dynamics along the time dimension, while also integrating the local structural features extracted by CNN. Experimental results confirm the critical role of LSTM in this task, serving as a bridge between short-term patterns and long-term trends within the overall CNN-LSTM architecture, and significantly improving predictive performance in terms of both MSE and RMSE metrics.

## 5. Model result analysis

The forecast curve in Figure 3 illustrates the predicted quarterly sales volume of a cancer drug in Egypt from 2015 to 2024 using the CNN-LSTM model. The x-axis represents time (quarters), and the y-axis shows the predicted sales volume. The blue curve indicates the actual sales data, while the red curve reflects the model's predicted values. Overall, the two curves closely align, suggesting high predictive accuracy. Notably, the model successfully captures seasonal fluctuations and long-term growth trends in sales volume. Minor deviations occur during periods of sudden market shifts, indicating that external shocks (e.g., policy changes or supply disruptions) may not be fully captured by the current model. Nonetheless, the forecast curve highlights the CNN-LSTM framework's strength in modeling nonlinear and volatile time-series data in the pharmaceutical field.

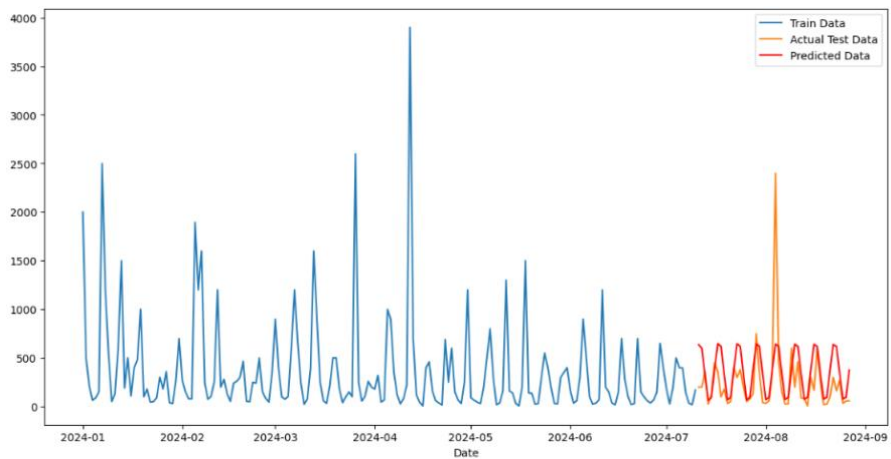

**Figure 3.** Forecast curve of sales volume of anti-cancer drugs

The CNN-LSTM-based model demonstrates strong capabilities in time series modeling and feature extraction for forecasting cancer drug sales volume [15]. The CNN component efficiently captures representative local temporal features in the sales data, such as price fluctuations, sales peaks, or periodic patterns, while the LSTM component excels at modeling long-term dependencies and overall trends over time. This synergy enables the model to effectively focus on short-term variations while also grasping long-term market dynamics, resulting in a robust and highly generalizable forecasting framework. In practical applications, this model offers valuable insights for pharmaceutical companies in production planning, inventory optimization, and proactive healthcare policy formulation.

**Table1.** Comparison of experimental results

| Model | MSE | RMSE |
| --- | --- | --- |
| CNN-LSTM | **1.150** | **1.072** |
| CNN | 3.526 | 1.878 |
| LSTM | 1.956 | 1.399 |
| RNN | 2.026 | 1.423 |

In the anticancer drug sales forecasting task, the CNN-LSTM model was compared with CNN, LSTM, and traditional RNN models. As shown in Table 1, the CNN-LSTM model achieved the lowest MSE of 1.150 and RMSE of 1.072. This was a significant improvement over the other models: CNN had an MSE of 3.526 and RMSE of 1.878, LSTM an MSE of 1.956 and RMSE of 1.399, and the traditional RNN an MSE of 2.026 and RMSE of 1.423.

The CNN-LSTM model's superior performance can be attributed to its unique architecture, which effectively captures complex features and long-term dependencies. The CNN layers extract local features from the data, while the LSTM layers model the long-term dependencies in the time-series data. This combination allows the CNN-LSTM model to more accurately capture market dynamics and underlying trends in complex time-series forecasting tasks compared to single models. Specifically, the CNN model's lack of effective long-term dependency modeling limits its forecasting accuracy. Traditional RNN models suffer from gradient vanishing issues, making it difficult to retain early information in long sequences and leading to poor forecasting performance. Although LSTM models excel at handling long-term dependencies, they are less effective at capturing local features.

In contrast, the CNN-LSTM model combines the strengths of CNN and LSTM to overcome these limitations. The CNN layers first extract features from the input data, capturing local patterns and structural information. Then, the LSTM layers model the extracted features over time, capturing long-term dependencies. This deep feature fusion enables the CNN-LSTM model to achieve better forecasting results by providing a more comprehensive understanding of the data.

## 5. Conclusions

This study focuses on modeling and forecasting cancer drug sales volume using time series data, proposing and validating a deep learning approach based on a CNN-LSTM framework. With the continuous rise in cancer incidence and advancements in oncology treatment technologies, the demand for cancer medications is steadily increasing in the global healthcare market. Accurate forecasting of cancer drug sales plays a crucial role in optimizing production planning, inventory management, and the formulation of healthcare policies. To achieve this, the study utilizes a quarterly dataset of a specific cancer drug sold in Egypt between 2015 and 2024, incorporating multidimensional features such as drug type, manufacturer, price, effectiveness, and more.

In the model architecture, the CNN component is responsible for extracting local temporal features from the sales sequence, identifying patterns such as sales peaks and price fluctuations. The LSTM component captures long-term dependencies, modeling seasonal trends and inter-quarter dynamics to effectively learn complex and nonlinear sales behaviors. A comprehensive data preprocessing pipeline—including missing value removal, feature normalization, and time alignment—was applied to ensure data quality and enhance model training effectiveness.

Experimental results demonstrate that the CNN-LSTM model performs with high accuracy in forecasting future cancer drug sales volumes. Evaluation on the test set yielded a Mean Squared Error (MSE) of 1.150 and a Root Mean Squared Error (RMSE) of 1.072, indicating strong generalization and stability in handling nonlinear and volatile medical sales data. The predicted sales trends closely align with actual observed values, confirming the model's ability to capture intricate market behaviors and generate reliable forecasts.

The findings of this research provide a data-driven decision-support tool for pharmaceutical companies and healthcare policy makers. The model can assist in ensuring timely drug availability for patients, reducing inventory overstock, and optimizing production resources. Additionally, it contributes theoretical and empirical insights into the application of deep learning in modeling complex medical-economic behavior.

Nonetheless, certain limitations remain. For instance, the model does not yet incorporate external variables such as policy changes, the launch of competing products, or unexpected public health events, all of which may introduce non-structural disruptions to sales trends. Furthermore, the current model is based on data from a single region and requires further validation to assess its robustness and adaptability in multi-regional or multi-drug scenarios.

Future research can explore several directions. First, incorporating macroeconomic indicators, policy documents, and patient behavior data may enrich the feature space and improve prediction accuracy. Second, advanced time-series modeling techniques such as attention mechanisms or Transformer-based architectures could be employed to enhance flexibility in long-sequence forecasting. Third, applying transfer learning and evaluating the model across diverse datasets could lead to the development of a scalable and generalizable cancer drug sales forecasting platform.